\documentclass[nofootinbib,prd]{revtex4}%
\usepackage{amsmath}
\usepackage{amsfonts}
\usepackage{amssymb}
\usepackage{graphicx}%
\def\beqa{\begin{eqnarray}}
\def\eeqa{\end{eqnarray}}
\def\beq{\begin{equation}}
\def\eeq{\end{equation}}

\renewcommand{\epsilon}{\varepsilon}

\let\in=\indent
\def\bet{\begin{tabular}}
\def\eet{\end{tabular}}
\def\bef{\begin{figure}}
\def\eef{\end{figure}}
\def\beqa{\begin{eqnarray}}
\def\eeqa{\end{eqnarray}}
\def\beq{\begin{equation}}
\def\eeq{\end{equation}}

\renewcommand{\epsilon}{\varepsilon}

\let\in=\indent
\def\beqa{\begin{eqnarray}}
\def\eeqa{\end{eqnarray}}
\def\beq{\begin{equation}}
\def\eeq{\end{equation}}

\renewcommand{\epsilon}{\varepsilon}

\begin{document}
\title{The cosmological constant as an eigenvalue of $f(R)$-gravity Hamiltonian constraint}
\author{S. Capozziello${}^{\sharp}$, R. Garattini${}^{\flat}$}
\thanks{\texttt{capozziello@na.infn.it; remo.garattini@unibg.it}}
\affiliation{${}^{\sharp}$ Dipartimento di Scienze Fisiche, Universit\`a di Napoli
"Federico II" and INFN Sez. di Napoli, Compl. Univ. Monte S. Angelo, Ed.N, Via
Cinthia, I-80126 Napoli, Italy}
\affiliation{${}^{\flat}$Facolt\`a di Ingegneria and INFN Sez. di Milano, Universit\`a di
Bergamo, Viale Marconi 5, I-24044 Dalmine (Bergamo), Italy}

\begin{abstract}
In the framework of ADM formalism, it is possible to find out eigenvalues of
the WDW equation with the meaning of vacuum states, i.e. cosmological
constants, for $f(R)$ theories of gravity, where $f(R)$ is a generic analytic
function of the Ricci curvature scalar $R$. The explicit calculation is
performed for a Schwarzschild metric where one-loop energy is derived by the
zeta function regularization method and a renormalized running $\Lambda_{0}$
constant is obtained.

\end{abstract}

\pacs{98.80.Cq, 98.80. Hw, 04.20.Jb, 04.50+h}
\maketitle

\section{Introduction}

General Relativity (GR), together with Quantum Field Theory, is
the major scientific achievement of last century. It is a theory
of spacetime, gravity and matter unifying these concepts in a
comprehensive scheme which gives rise to a new conception of the
Universe. However, in the last thirty years, several shortcomings
came out in the Einstein scheme and people began to investigate if
GR is the only theory able to explain the gravitational
interaction. Such issues essentially spring up in Cosmology and
Quantum Field Theory. In the first case, the presence of Big Bang
singularity, flatness and horizon problems \cite{guth} led to the
result that Standard Cosmological Model \cite{weinberg}, is
inadequate to describe the Universe at extreme regimes. On the
other hand, GR is a \textit{classical} theory which does not work
as a fundamental theory, when one wants to achieve a full quantum
description of spacetime (and then of gravity). Due to this facts
and, first of all, to the lack of a definitive Quantum Gravity
theory, alternative theories of gravity have been pursued in order
to attempt, at least, a semi-classical scheme where GR and its
positive results could be recovered. A fruitful approach has been
that of \textit{Extended Theories of Gravity} (ETG) which have
become a sort of paradigm in the study of gravitational
interaction based on corrections and enlargements of the Einstein
scheme. The paradigm consists, essentially, in adding higher-order
curvature invariants and non-minimally coupled scalar fields into
dynamics resulting from the effective action of Quantum Gravity
\cite{odintsov,farhoudi}.

All these approaches are not the ``\textit{full Quantum Gravity}"
but are needed as working schemes toward it. In any case, they are
going to furnish consistent and physically reliable results.
Furthermore, every unification scheme as Superstrings,
Supergravity or Grand Unified Theories, takes into account
effective actions where non-minimal couplings to the geometry or
higher-order terms in the curvature invariants come out. Such
contributions are due to one-loop or higher-loop corrections in
the high-curvature regimes. Specifically, this scheme has been
adopted in order to deal with quantization on curved spacetimes
and the result has been that the interactions among quantum scalar
fields and background geometry or the gravitational
self-interactions yield corrective terms in the Hilbert-Einstein
Lagrangian \cite{birrell}. Moreover, it has been realized that
such corrective terms are inescapable if we want to obtain the
effective action of Quantum Gravity on scales closed to the Planck
length \cite{vilkovisky}.

Besides fundamental physics motivations, all these theories have
acquired a huge interest in cosmology due to the fact that they
``naturally" exhibit inflationary behaviors able to overcome the
shortcomings of Standard Cosmological Model (based on GR). The
related cosmological models seem very realistic and, several
times, capable of matching with the observations
\cite{starobinsky,la,kerner}. Furthermore, it is possible to show
that, via conformal transformations, the higher-order and
non-minimally coupled terms always correspond to Einstein gravity
plus one or more than one minimally coupled scalar fields
\cite{teyssandier,maeda,wands,gottloeber}.  This feature results
very interesting if we want to obtain multiple inflationary events
since a former early stage could select ``very'' large-scale
structures (clusters of galaxies today), while a latter stage
could select ``small'' large-scale structures (galaxies today)
\cite{sixth}. The philosophy is that each inflationary era is
connected with the dynamics of a scalar field. Furthermore, these
extended schemes naturally could solve the problem of ``graceful
exit" bypassing the shortcomings of former inflationary models
\cite{la,aclo}.

Recently, ETG are going also to play an interesting role to
describe the today observed Universe. In fact, the amount of good
quality data of last decade has made it possible to shed new light
on the effective picture of the Universe. Type Ia Supernovae
(SNeIa) \cite{SNeIa}, anisotropies in the cosmic microwave
background radiation (CMBR) \cite{CMBR}, and matter power spectrum
inferred from large galaxy surveys \cite{LSS} represent the
strongest evidences for a radical revision of the Cosmological
Standard Model also at recent epochs. In particular, the
\textit{concordance $\Lambda$CDM model} predicts that baryons
contribute only for $\sim4\%$ of the total matter\,-\,energy
budget, while the exotic \textit{cold dark matter} (CDM)
represents the bulk of the matter content ($\sim25\%$) and the
cosmological constant $\Lambda$ plays the role of the so called
"dark energy" ($\sim70\%$) \cite{triangle}. Although being the
best fit to a wide range of data \cite{LambdaTest}, the
$\Lambda$CDM model is severely affected by strong theoretical
shortcomings \cite{LambdaRev} that have motivated the search for
alternative models \cite{PR03}. Dark energy models mainly rely on
the implicit assumption that Einstein's General Relativity is the
correct theory of gravity indeed. Nevertheless, its validity on
the larger astrophysical and cosmological scales has never been
tested \cite{will}, and it is therefore conceivable that both
cosmic speed up and dark matter represent signals of a breakdown
in our understanding of gravitation law so that one should
consider the possibility that the Hilbert\,-\,Einstein Lagrangian,
linear in the Ricci scalar $R$, should be generalized. Following
this line of thinking, the choice of a generic function $f(R)$ can
be derived by matching the data and by the "economic" requirement
that no exotic ingredients have to be added\footnote{Following the
Occam razor prescriptions: \textit{"Entia non sunt multiplicanda
praeter necessitatem."}}. This is the underlying philosophy of
what are referred to as $f(R)$ theories of gravity, see
\cite{capozzcurv,curvrev,cdtt,flanagan,allemandi,odintsovfr,mimicking}
and references therein. However $f(R)$ gravity can be encompassed
in the ETG being a "minimal" extension of GR where (analytical)
functions of Ricci scalar are taken into account. Although higher
order gravity theories have received much attention in cosmology,
since they are naturally able to give rise to the accelerating
expansion (both in the late and in the early universe, it is
possible to demonstrate that $f(R)$ theories can also play a major
role at astrophysical scales. In fact, modifying the gravity
Lagrangian can affect the gravitational potential in the low
energy limit. Provided that the modified potential reduces to the
Newtonian one on the Solar System scale, this implication could
represent an intriguing opportunity rather than a shortcoming for
$f(R)$ theories. In fact, a corrected gravitational potential
could  fit galaxy rotation curves without the need of dark matter
\cite{noipla,mond,jcap}. In addition, it is possible to work out a
formal analogy between the corrections to the Newtonian potential
and the usually adopted dark matter models. In general, any
relativistic theory of gravitation can yield corrections to the
Newton potential (see for example \cite{schmidt}) which, in the
post-Newtonian (PPN) formalism, could furnish tests for the same
theory \cite{will,ppnantro,arturo,matteo}.

In this paper, we want to face the problem to study $f(R)$ gravity
at a fundamental level. In particular, in the framework of the
Arnowitt-Deser-Misner ($\mathcal{ADM}$) formalism \cite{ADM}, we
want to investigate the possibility to find out cosmological terms
as eigenvalues of generalized $f(R)$-Hamiltonians in a
Sturm-Liouville-like problem\footnote{See Ref.\cite{Remo1}, for
the application of the Sturm-Liouville problem in the simple case
of $f\left(  R\right)  =R$, even in presence of a massive
graviton.}. This issue is particularly relevant from several
viewpoints. First of all, our aim is to show that vacuum energy of
gravitational field is not a particular feature of GR where the
cosmological constant has to by added \textit{by hand} into
dynamics. At a classical level, it is well known that $f(R)$
gravity, for the Ricci scalar $R$ equal to a constant, exhibit
several deSitter solutions \cite{ottewill} but a definite
discussion, at a fundamental level, considering cosmological terms
as eigenvalues of such theories is lacking. Besides, the
computation of the Casimir energy, the seeking for zero point
energy in different backgrounds\footnote{For different $f(R)$, we
expect different zero point energies and, obviously, different
vacuum states.} give a track to achieve one-loop energy
regularization and renormalization for this kind of theories
\cite{zerbini,zerbini2}. On the other hand, these issues can be
considered in a multigravity approach to spacetime foam if the
$\mathcal{N}$ spacetimes constituting the foam are supposed to
evolve, in general, with different curvature laws and ground
states (cosmological constants) \cite{Remo}.

The layout of the paper is the following. In Sec.II, we recall the Hamiltonian
formalism in the ($\mathcal{ADM}$) approach of GR. It is developed for generic
$f(R)$ gravity in Sec.III. Sec.IV is devoted to find out the cosmological
constant as the eigenvalue of a generalized $f(R)$ Hamiltonian. We discuss the
orthogonal decomposition of the wave functional and derive the total one-loop
energy density for the transverse-traceless tensor component. In Sec.V, we
give an example: the transverse-traceless spin 2 operator is calculated for
the Schwarzschild metric and the energy density contributions to the
cosmological constant are calculated in the WKB approximation. This is a
realization of the above formal cosmological constant calculation. Sec.VI is
devoted to the one-loop energy calculation by the zeta function regularization
method. The explicit value of the renormalized $\Lambda_{0}$ constant,
considered as a running constant, is achieved. How it can be set to zero is
explicitly derived for $f(R)=\exp(-\alpha R)$. Summary and conclusions are
drawn in Sec. VII. In Appendix A, details on zeta function regularization are given.

\section{The Hamiltonian constraint of General Relativity}

Let us briefly report how to compute the Hamiltonian constraint for GR
considering the standard Hilbert-Einstein theory $f\left(  R\right)  =R$ and
the Arnowitt-Deser-Misner ($\mathcal{ADM}$) $3+1$ decomposition \cite{ADM}. In
terms of these variables, the line element is
\[
ds^{2}=g_{\mu\nu}\left(  x\right)  dx^{\mu}dx^{\nu}=\left(  -N^{2}+N_{i}%
N^{i}\right)  dt^{2}+2N_{j}dtdx^{j}+g_{ij}dx^{i}dx^{j}.
\]
$N$ is the \textit{lapse function}, while $N_{i}$ the \textit{shift function}.
In terms of these variables, the gravitational Lagrangian, with the boundary
terms neglected, can be written as
\begin{equation}
\mathcal{L}\left[  N,N_{i},g_{ij}\right]  =\sqrt{-g}R=\frac{N{}\,\sqrt{^{3}g}%
}{2\kappa}\text{ }\left[  K_{ij}K^{ij}-K^{2}+\,\left(  ^{3}R-2\Lambda
_{c}\right)  \right]  ,
\end{equation}
where $K_{ij}$ is the second fundamental form, $K=$ $g^{ij}K_{ij}$ is the
trace, $^{3}R$ is the three dimensional scalar curvature and $\sqrt{^{3}g}$ is
the three dimensional determinant of the metric. The conjugate momentum is
simply%
\begin{equation}
\pi^{ij}=\frac{\delta\mathcal{L}}{\delta\left(  \partial_{t}g_{ij}\right)
}=\left(  ^{3}g^{ij}K-K^{ij}\text{ }\right)  \frac{\sqrt{^{3}g}}{2\kappa
}.\label{mom}%
\end{equation}
By a Legendre transformation, we calculate the Hamiltonian
\begin{equation}
H=%
%TCIMACRO{\dint }%
%BeginExpansion
{\displaystyle\int}
%EndExpansion
d^{3}x\left[  N\mathcal{H+}N_{i}\mathcal{H}^{i}\right]  ,
\end{equation}
where%
\begin{equation}
\mathcal{H}=\left(  2\kappa\right)  G_{ijkl}\pi^{ij}\pi^{kl}-\frac{\sqrt
{^{3}g}}{2\kappa}\left(  ^{3}R-2\Lambda_{c}\right)  \label{cla1}%
\end{equation}
and%
\begin{equation}
\mathcal{H}^{i}=-2\nabla_{j}\pi^{ji}.\label{cla2}%
\end{equation}
where $\Lambda_{c}$ is the bare cosmological constant. The equations of motion
lead to two classical constraints%
\begin{equation}
\left\{
\begin{array}
[c]{c}%
\mathcal{H}=0\\
\mathcal{H}^{i}=0
\end{array}
\right.  ,
\end{equation}
representing invariance under \textit{time} re-parameterization and invariance
under diffeomorphism, respectively. $G_{ijkl}$ is the \textit{supermetric}
defined as%
\begin{equation}
G_{ijkl}=\frac{1}{2\sqrt{g}}(g_{ik}g_{jl}+g_{il}g_{jk}-g_{ij}g_{kl}).
\end{equation}
When $\mathcal{H}$ and $\mathcal{H}^{i}$ are considered as operators acting on
some wave function, we have%
\begin{equation}
\mathcal{H}\Psi\left[  g_{ij}\right]  =0\label{WDW}%
\end{equation}
and%
\begin{equation}
\mathcal{H}_{i}\Psi\left[  g_{ij}\right]  =0.\label{diff}%
\end{equation}
Eq.$\left(  \ref{WDW}\right)  $ is the Wheeler-De Witt equation (WDW)\cite{De
Witt}. Eqs.$\left(  \ref{WDW}\right)  $ and $\left(  \ref{diff}\right)  $
describe the \textit{wave function of the universe }$\Psi\left[
g_{ij}\right]  $. The WDW equation represents invariance under \textit{time}
re-parameterization in an operatorial form. This standard lore can be applied
to a generic $f(R)$ theory of gravity with the aim to achieve a cosmological
term as an eigenvalue of the WDW equation.

\section{The Hamiltonian constraint for a generic $f(R)$ theory of gravity}

Let us consider now the Lagrangian density describing a generic $f(R)$ theory
of gravity, namely
\begin{equation}
\mathcal{L}=\sqrt{-g}\left(  f\left(  R\right)  -2\Lambda_{c}\right)
,\qquad\text{with}\;f^{\prime\prime}\neq0,\label{lag}%
\end{equation}
where $f\left(  R\right)  $ is an arbitrary smooth function of the scalar
curvature and primes denote differentiation with respect to the scalar
curvature. A cosmological term is added also in this case for the sake of
generality. Obviously $f^{\prime\prime}=0$ corresponds to GR. The generalized
Hamiltonian density for the $f\left(  R\right)  $ theory assumes the
form\footnote{See also Ref.\cite{Querella} for technical details.}%
\begin{equation}
\mathcal{H}=\frac{1}{2\kappa}\left[  \frac{\mathcal{P}}{6}\left(  {}^{\left(
3\right)  }R-2\Lambda_{c}-3K_{ij}K^{ij}+K^{2}\right)  +V(\mathcal{P})-\frac
{1}{3}g^{ij}\mathcal{P}_{\mid ij}-2p^{ij}K_{ij}\right]  ,
\end{equation}
where%
\begin{equation}
V(\mathcal{P})=\sqrt{g}\left[  Rf^{\prime}\left(  R\right)  -f\left(
R\right)  \right]  .\label{V(P)}%
\end{equation}
Henceforth, the superscript $3$ indicating the spatial part of the metric will
be omitted on the metric itself. When $f\left(  R\right)  =R$, $V(\mathcal{P}%
)=0$ as it should be. Since%
\begin{equation}
\mathcal{P}^{ij}=-2\sqrt{g}g^{ij}f^{\prime}\left(  R\right)  \qquad
\Longrightarrow\qquad\mathcal{P=}-6\sqrt{g}f^{\prime}\left(  R\right)  ,
\end{equation}
we have%
\begin{equation}
\mathcal{H}=\frac{1}{2\kappa}\left[  -\sqrt{g}f^{\prime}\left(  R\right)
\left(  {}^{\left(  3\right)  }R-2\Lambda_{c}-3K_{ij}K^{ij}+K^{2}\right)
+V(\mathcal{P})+2g^{ij}\left(  \sqrt{g}f^{\prime}\left(  R\right)  \right)
_{\mid ij}-2p^{ij}K_{ij}\right]  .\label{Hamf(R)}%
\end{equation}
With the help of Eq.$\left(  \ref{mom}\right)  $, Eq.$\left(  \ref{Hamf(R)}%
\right)  $ becomes%
\begin{equation}
\mathcal{H}=f^{\prime}\left(  R\right)  \left[  \left(  2\kappa\right)
G_{ijkl}\pi^{ij}\pi^{kl}{}-\frac{\sqrt{g}}{2\kappa}{}\left(  ^{\left(
3\right)  }R-2\Lambda_{c}\right)  \right]  +\frac{1}{2\kappa}\left[  \sqrt
{g}f^{\prime}\left(  R\right)  \left(  2K_{ij}K^{ij}\right)  +V(\mathcal{P}%
)+2g^{ij}\left(  \sqrt{g}f^{\prime}\left(  R\right)  \right)  _{\mid
ij}-2p^{ij}K_{ij}\right]  .
\end{equation}
However%
\begin{equation}
p^{ij}=\sqrt{g}K^{ij},
\end{equation}
then we obtain%
\begin{equation}
\mathcal{H}=f^{\prime}\left(  R\right)  \left[  \left(  2\kappa\right)
G_{ijkl}\pi^{ij}\pi^{kl}{}-\frac{\sqrt{g}}{2\kappa}{}\left(  ^{\left(
3\right)  }R-2\Lambda_{c}\right)  \right]  +\frac{1}{2\kappa}\left[  2\sqrt
{g}K_{ij}K^{ij}\left(  f^{\prime}\left(  R\right)  -1\right)  +V(\mathcal{P}%
)+2g^{ij}\left(  \sqrt{g}f^{\prime}\left(  R\right)  \right)  _{\mid
ij}\right]
\end{equation}
and transforming into canonical momenta, one gets
\begin{equation}
\mathcal{H}=f^{\prime}\left(  R\right)  \left[  \left(  2\kappa\right)
G_{ijkl}\pi^{ij}\pi^{kl}{}-\frac{\sqrt{g}}{2\kappa}{}\left(  ^{\left(
3\right)  }R-2\Lambda_{c}\right)  \right]  +2\left(  2\kappa\right)  \left[
G_{ijkl}\pi^{ij}\pi^{kl}+\frac{\pi}{4}^{2}\right]  \left(  f^{\prime}\left(
R\right)  -1\right)  +\frac{1}{2\kappa}\left[  V(\mathcal{P})+2g^{ij}\left(
\sqrt{g}f^{\prime}\left(  R\right)  \right)  _{\mid ij}\right]
.\label{Hamf(R)_1}%
\end{equation}
By imposing the Hamiltonian constraint, we obtain
\begin{equation}
f^{\prime}\left(  R\right)  \left[  \left(  2\kappa\right)  G_{ijkl}\pi
^{ij}\pi^{kl}{}-\frac{\sqrt{g}}{2\kappa}{}^{\left(  3\right)  }R\right]  +{}%
\end{equation}%
\begin{equation}
+2\left(  2\kappa\right)  \left[  G_{ijkl}\pi^{ij}\pi^{kl}+\frac{\pi}{4}%
^{2}\right]  \left(  f^{\prime}\left(  R\right)  -1\right)  +\frac{1}{2\kappa
}\left[  V(\mathcal{P})+2g^{ij}\left(  \sqrt{g}f^{\prime}\left(  R\right)
\right)  _{\mid ij}\right]  =-f^{\prime}\left(  R\right)  \sqrt{g}%
\frac{\Lambda_{c}}{\kappa}%
\end{equation}
If we assume that $f^{\prime}\left(  R\right)  \neq0$ the previous expression
becomes%
\begin{equation}
\left[  \left(  2\kappa\right)  G_{ijkl}\pi^{ij}\pi^{kl}{}-\frac{\sqrt{g}%
}{2\kappa}{}^{\left(  3\right)  }R\right]  +\left(  2\kappa\right)  \left[
G_{ijkl}\pi^{ij}\pi^{kl}+\frac{\pi}{4}^{2}\right]  \frac{2\left(  f^{\prime
}\left(  R\right)  -1\right)  }{f^{\prime}\left(  R\right)  }+\frac{1}{2\kappa
f^{\prime}\left(  R\right)  }\left[  V(\mathcal{P})+2g^{ij}\left(  \sqrt
{g}f^{\prime}\left(  R\right)  \right)  _{\mid ij}\right]  =-\sqrt{g}%
\frac{\Lambda_{c}}{\kappa}{}.
\end{equation}
Now, we integrate over the hypersurface $\Sigma$ to obtain%
\begin{equation}
\int_{\Sigma}d^{3}x\left\{  \left[  \left(  2\kappa\right)  G_{ijkl}\pi
^{ij}\pi^{kl}{}-\frac{\sqrt{g}}{2\kappa}{}^{\left(  3\right)  }R\right]
+\left(  2\kappa\right)  \left[  G_{ijkl}\pi^{ij}\pi^{kl}+\frac{\pi}{4}%
^{2}\right]  \frac{2\left(  f^{\prime}\left(  R\right)  -1\right)  }%
{f^{\prime}\left(  R\right)  }\right\}
\end{equation}%
\begin{equation}
+\int_{\Sigma}d^{3}x\frac{1}{2\kappa f^{\prime}\left(  R\right)  }\left[
V(\mathcal{P})+2g^{ij}\left(  \sqrt{g}f^{\prime}\left(  R\right)  \right)
_{\mid ij}\right]  =-\frac{\Lambda_{c}}{\kappa}\int_{\Sigma}d^{3}x\sqrt{g}.
\end{equation}
The term%
\begin{equation}
\frac{1}{\kappa}\int_{\Sigma}d^{3}x\frac{1}{f^{\prime}\left(  R\right)
}g^{ij}\left(  \sqrt{g}f^{\prime}\left(  R\right)  \right)  _{\mid ij}%
\end{equation}
appears to be a three-divergence and therefore will not contribute to the
computation. The remaining equation simplifies into%
\begin{equation}
\int_{\Sigma}d^{3}x\left\{  \left[  \left(  2\kappa\right)  G_{ijkl}\pi
^{ij}\pi^{kl}{}-\frac{\sqrt{g}}{2\kappa}{}^{\left(  3\right)  }R\right]
+\left(  2\kappa\right)  \left[  G_{ijkl}\pi^{ij}\pi^{kl}+\frac{\pi}{4}%
^{2}\right]  \frac{2\left(  f^{\prime}\left(  R\right)  -1\right)  }%
{f^{\prime}\left(  R\right)  }+\frac{V(\mathcal{P})}{2\kappa f^{\prime}\left(
R\right)  }\right\}  =-\frac{\Lambda_{c}}{\kappa}\int_{\Sigma}d^{3}x\sqrt
{g}.\label{GWDW}%
\end{equation}
By a canonical procedure of quantization, we want to obtain the vacuum state
of a generic $f(R)$ theory.

\section{The cosmological constant as an eigenvalue for the generalized
$f\left(  R\right)  $ Hamiltonian}

The standard WDW equation $\left(  \ref{WDW}\right)  $ can be cast into the
form of an eigenvalue equation
\begin{equation}
\hat{\Lambda}_{\Sigma}\Psi\left[  g_{ij}\right]  =\Lambda\left(  \vec
{x}\right)  \Psi\left[  g_{ij}\right]  , \label{WDW1}%
\end{equation}
where
\begin{equation}
\hat{\Lambda}_{\Sigma}=\left(  2\kappa\right)  G_{ijkl}\pi^{ij}\pi^{kl}%
-\frac{\sqrt{g}}{2\kappa}\!^{3}R.
\end{equation}
If we multiply Eq.$\left(  \ref{WDW1}\right)  $ by $\Psi^{\ast}\left[
g_{ij}\right]  $ and we functionally integrate over the three spatial metric
$g_{ij}$, we get%
\begin{equation}
\int\mathcal{D}\left[  g_{ij}\right]  \Psi^{\ast}\left[  g_{ij}\right]
\hat{\Lambda}_{\Sigma}\Psi\left[  g_{ij}\right]  =\int\mathcal{D}\left[
g_{ij}\right]  \Lambda\left(  \vec{x}\right)  \Psi^{\ast}\left[
g_{ij}\right]  \Psi\left[  g_{ij}\right]
\end{equation}
and after integrating over the hypersurface $\Sigma$, one can formally
re-write the modified WDW equation as%
\begin{equation}
\frac{1}{V}\frac{\int\mathcal{D}\left[  g_{ij}\right]  \Psi^{\ast}\left[
g_{ij}\right]  \int_{\Sigma}d^{3}x\hat{\Lambda}_{\Sigma}\Psi\left[
g_{ij}\right]  }{\int\mathcal{D}\left[  g_{ij}\right]  \Psi^{\ast}\left[
g_{ij}\right]  \Psi\left[  g_{ij}\right]  }=\frac{1}{V}\frac{\left\langle
\Psi\left\vert \int_{\Sigma}d^{3}x\hat{\Lambda}_{\Sigma}\right\vert
\Psi\right\rangle }{\left\langle \Psi|\Psi\right\rangle }=-\frac{\Lambda_{c}%
}{\kappa}, \label{WDW2}%
\end{equation}
where the explicit expression of $\Lambda\left(  \vec{x}\right)  $ has been
used and we have defined the volume of the hypersurface $\Sigma$ as%
\begin{equation}
V=\int_{\Sigma}d^{3}x\sqrt{g}.
\end{equation}
The formal eigenvalue equation $\left(  \ref{WDW2}\right)  $ is a simple
manipulation of Eq.$\left(  \ref{WDW}\right)  $. We can gain more information
considering a separation of the spatial part of the metric into a background
term, $\bar{g}_{ij}$, and a quantum fluctuation, $h_{ij}$,%
\begin{equation}
g_{ij}=\bar{g}_{ij}+h_{ij}.
\end{equation}
Thus Eq.$\left(  \ref{WDW2}\right)  $ becomes%
\begin{equation}
\frac{\left\langle \Psi\left\vert \int_{\Sigma}d^{3}x\left[  \hat{\Lambda
}_{\Sigma}^{\left(  0\right)  }+\hat{\Lambda}_{\Sigma}^{\left(  1\right)
}+\hat{\Lambda}_{\Sigma}^{\left(  2\right)  }+\ldots\right]  \right\vert
\Psi\right\rangle }{\left\langle \Psi|\Psi\right\rangle }=-\frac{\Lambda_{c}%
}{\kappa}\Psi\left[  g_{ij}\right]  , \label{WDW3}%
\end{equation}
where $\hat{\Lambda}_{\Sigma}^{\left(  i\right)  }$ represents the $i^{th}$
order of perturbation in $h_{ij}$. By observing that the kinetic part of
$\hat{\Lambda}_{\Sigma}$ is quadratic in the momenta, we only need to expand
the three-scalar curvature $\int d^{3}x\sqrt{g}R^{\left(  3\right)  }$ up to
the quadratic order and we get%
\[
\int_{\Sigma}d^{3}x\sqrt{\bar{g}}\left[  -\frac{1}{4}h\triangle h+\frac{1}%
{4}h^{li}\triangle h_{li}-\frac{1}{2}h^{ij}\nabla_{l}\nabla_{i}h_{j}%
^{l}+\right.
\]%
\begin{equation}
\left.  +\frac{1}{2}h\nabla_{l}\nabla_{i}h^{li}-\frac{1}{2}h^{ij}R_{ia}%
h_{j}^{a}+\frac{1}{2}hR_{ij}h^{ij}+\frac{1}{4}h\left(  R^{\left(  0\right)
}\right)  h\right]  \label{rexp}%
\end{equation}
where $h$ is the trace of $h_{ij}$ and $R^{\left(  0\right)  }$ is the three
dimensional scalar curvature. By repeating the same procedure for the
generalized WDW equation Eq.$\left(  \ref{GWDW}\right)  $, we obtain%
\begin{equation}
\frac{1}{V}\frac{\left\langle \Psi\left\vert \int_{\Sigma}d^{3}x\left[
\hat{\Lambda}_{\Sigma}^{\left(  2\right)  }\right]  \right\vert \Psi
\right\rangle }{\left\langle \Psi|\Psi\right\rangle }+\frac{2\kappa}{V}%
\frac{2\left(  f^{\prime}\left(  R\right)  -1\right)  }{f^{\prime}\left(
R\right)  }\frac{\left\langle \Psi\left\vert \int_{\Sigma}d^{3}x\left[
G_{ijkl}\pi^{ij}\pi^{kl}+\frac{\pi}{4}^{2}\right]  \right\vert \Psi
\right\rangle }{\left\langle \Psi|\Psi\right\rangle }+\frac{1}{V}%
\frac{\left\langle \Psi\left\vert \int_{\Sigma}d^{3}x\frac{V(\mathcal{P}%
)}{2\kappa f^{\prime}\left(  R\right)  }\right\vert \Psi\right\rangle
}{\left\langle \Psi|\Psi\right\rangle }=-\frac{\Lambda_{c}}{\kappa}.
\label{GWDW1}%
\end{equation}
From Eq.$\left(  \ref{GWDW1}\right)  $, we can define a \textquotedblleft%
\textit{modified}\textquotedblright\ $\hat{\Lambda}_{\Sigma}^{\left(
2\right)  }$ operator which includes $f^{\prime}\left(  R\right)  $. Thus, we
obtain%
\begin{equation}
\frac{\left\langle \Psi\left\vert \int_{\Sigma}d^{3}x\left[  \hat{\Lambda
}_{\Sigma,f\left(  R\right)  }^{\left(  2\right)  }\right]  \right\vert
\Psi\right\rangle }{\left\langle \Psi|\Psi\right\rangle }+\frac{2\kappa}%
{V}\frac{2\left(  f^{\prime}\left(  R\right)  -1\right)  }{f^{\prime}\left(
R\right)  }\frac{\left\langle \Psi\left\vert \int_{\Sigma}d^{3}x\left[
\frac{\pi}{4}^{2}\right]  \right\vert \Psi\right\rangle }{\left\langle
\Psi|\Psi\right\rangle }+\frac{1}{V}\frac{\left\langle \Psi\left\vert
\int_{\Sigma}d^{3}x\frac{V(\mathcal{P})}{2\kappa f^{\prime}\left(  R\right)
}\right\vert \Psi\right\rangle }{\left\langle \Psi|\Psi\right\rangle }%
=-\frac{\Lambda_{c}}{\kappa}, \label{GWDW2}%
\end{equation}
where%
\begin{equation}
\hat{\Lambda}_{\Sigma,f\left(  R\right)  }^{\left(  2\right)  }=\left(
2\kappa\right)  h\left(  R\right)  G_{ijkl}\pi^{ij}\pi^{kl}-\frac{\sqrt{g}%
}{2\kappa}\text{ }^{3}R^{lin},
\end{equation}
with%
\begin{equation}
h\left(  R\right)  =1+\frac{2\left[  f^{\prime}\left(  R\right)  -1\right]
}{f^{\prime}\left(  R\right)  } \label{h(R)}%
\end{equation}
and where $^{3}R^{lin}$ is the linearized scalar curvature whose expression is
shown in square brackets of Eq.$\left(  \ref{rexp}\right)  $. Note that when
$f\left(  R\right)  =R$, consistently it is $h\left(  R\right)  =1$. From
Eq.$\left(  \ref{GWDW2}\right)  $, we redefine $\Lambda_{c}$%
\begin{equation}
\Lambda_{c}^{\prime}=\Lambda_{c}+\frac{1}{2V}\frac{\left\langle \Psi\left\vert
\int_{\Sigma}d^{3}x\frac{V(\mathcal{P})}{f^{\prime}\left(  R\right)
}\right\vert \Psi\right\rangle }{\left\langle \Psi|\Psi\right\rangle }%
=\Lambda_{c}+\frac{1}{2V}\int_{\Sigma}d^{3}x\sqrt{g}\frac{Rf^{\prime}\left(
R\right)  -f\left(  R\right)  }{f^{\prime}\left(  R\right)  },
\label{NewLambda}%
\end{equation}
where we have explicitly used the definition of $V(\mathcal{P})$. In order to
make explicit calculations, we need an orthogonal decomposition for both
$\pi_{ij\text{ }}$ and $h_{ij}$ to disentangle gauge modes from physical
deformations. We define the inner product%

\begin{equation}
\left\langle h,k\right\rangle :=\int_{\Sigma}\sqrt{g}G^{ijkl}h_{ij}\left(
x\right)  k_{kl}\left(  x\right)  d^{3}x,
\end{equation}
by means of the inverse WDW metric $G_{ijkl}$, to have a metric on the space
of deformations, i.e. a quadratic form on the tangent space at $h_{ij}$, with%

\begin{equation}%
\begin{array}
[c]{c}%
G^{ijkl}=\frac{1}{2}(g^{ik}g^{jl}+g^{il}g^{jk}-2g^{ij}g^{kl})\text{.}%
\end{array}
\end{equation}
The inverse metric is defined on cotangent space and it assumes the form%

\begin{equation}
\left\langle p,q\right\rangle :=\int_{\Sigma}\sqrt{g}G_{ijkl}p^{ij}\left(
x\right)  q^{kl}\left(  x\right)  d^{3}x,
\end{equation}
so that%

\begin{equation}
G^{ijnm}G_{nmkl}=\frac{1}{2}\left(  \delta_{k}^{i}\delta_{l}^{j}+\delta
_{l}^{i}\delta_{k}^{j}\right)  .
\end{equation}
Note that in this scheme the \textquotedblleft inverse
metric\textquotedblright\ is actually the WDW metric defined on phase space.
The desired decomposition on the tangent space of 3-metric
deformations\cite{BergerEbin,York,MazurMottola,Vassilevich} is:%

\begin{equation}
h_{ij}=\frac{1}{3}hg_{ij}+\left(  L\xi\right)  _{ij}+h_{ij}^{\bot}
\label{p21a}%
\end{equation}
where the operator $L$ maps $\xi_{i}$ into symmetric tracefree tensors%

\begin{equation}
\left(  L\xi\right)  _{ij}=\nabla_{i}\xi_{j}+\nabla_{j}\xi_{i}-\frac{2}%
{3}g_{ij}\left(  \nabla\cdot\xi\right)  .
\end{equation}
Thus the inner product between three-geometries becomes
\[
\left\langle h,h\right\rangle :=\int_{\Sigma}\sqrt{g}G^{ijkl}h_{ij}\left(
x\right)  h_{kl}\left(  x\right)  d^{3}x=
\]%
\begin{equation}
\int_{\Sigma}\sqrt{g}\left[  -\frac{2}{3}h^{2}+\left(  L\xi\right)
^{ij}\left(  L\xi\right)  _{ij}+h^{ij\bot}h_{ij}^{\bot}\right]  . \label{p21b}%
\end{equation}
With the orthogonal decomposition in hand we can define the trial wave
functional as%
\begin{equation}
\Psi\left[  h_{ij}\left(  \overrightarrow{x}\right)  \right]  =\mathcal{N}%
\Psi\left[  h_{ij}^{\bot}\left(  \overrightarrow{x}\right)  \right]
\Psi\left[  h_{ij}^{\Vert}\left(  \overrightarrow{x}\right)  \right]
\Psi\left[  h_{ij}^{trace}\left(  \overrightarrow{x}\right)  \right]  ,
\label{twf}%
\end{equation}
where
\begin{equation}%
\begin{array}
[c]{c}%
\Psi\left[  h_{ij}^{\bot}\left(  \overrightarrow{x}\right)  \right]
=\exp\left\{  -\frac{1}{4}\left\langle hK^{-1}h\right\rangle _{x,y}^{\bot
}\right\} \\
\\
\Psi\left[  h_{ij}^{\Vert}\left(  \overrightarrow{x}\right)  \right]
=\exp\left\{  -\frac{1}{4}\left\langle \left(  L\xi\right)  K^{-1}\left(
L\xi\right)  \right\rangle _{x,y}^{\Vert}\right\} \\
\\
\Psi\left[  h_{ij}^{trace}\left(  \overrightarrow{x}\right)  \right]
=\exp\left\{  -\frac{1}{4}\left\langle hK^{-1}h\right\rangle _{x,y}%
^{Trace}\right\}
\end{array}
.
\end{equation}
The symbol \textquotedblleft$\perp$\textquotedblright\ denotes the
transverse-traceless tensor (TT) (spin 2) of the perturbation, while the
symbol \textquotedblleft$\Vert$\textquotedblright\ denotes the longitudinal
part (spin 1) of the perturbation. Finally, the symbol \textquotedblleft%
$trace$\textquotedblright\ denotes the scalar part of the perturbation.
$\mathcal{N}$ is a normalization factor, $\left\langle \cdot,\cdot
\right\rangle _{x,y}$ denotes space integration and $K^{-1}$ is the inverse
\textquotedblleft\textit{propagator}\textquotedblright. We will fix our
attention to the TT tensor sector of the perturbation representing the
graviton. Therefore, representation $\left(  \ref{twf}\right)  $ reduces to%
\begin{equation}
\Psi\left[  h_{ij}\left(  \overrightarrow{x}\right)  \right]  =\mathcal{N}%
\exp\left\{  -\frac{1}{4}\left\langle hK^{-1}h\right\rangle _{x,y}^{\bot
}\right\}  . \label{tt}%
\end{equation}
Actually there is no reason to neglect longitudinal and trace perturbations.
However, following the analysis of Refs.\cite{MazurMottola,GPY,VolkovWipf} on
the perturbation decomposition, we can discover that the relevant components
can be restricted to the TT modes and to the trace modes. Moreover, for
certain backgrounds, TT tensors can be a source of instability as shown in
Refs.\cite{GPY,VolkovWipf,Instability}. Even the trace part can be regarded as
a source of instability. Indeed this is usually termed \textit{conformal
}instability. The appearance of an instability on the TT modes is known as
non-conformal instability. This means that does not exist a gauge choice that
can eliminate negative modes. To proceed with Eq.$\left(  \ref{WDW3}\right)
$, we need to know the action of some basic operators on $\Psi\left[
h_{ij}\right]  $. The action of the operator $h_{ij}$ on $|\Psi\rangle
=\Psi\left[  h_{ij}\right]  $ is realized by \cite{Variational}
\begin{equation}
h_{ij}\left(  x\right)  |\Psi\rangle=h_{ij}\left(  \overrightarrow{x}\right)
\Psi\left[  h_{ij}\right]  .
\end{equation}
The action of the operator $\pi_{ij}$ on $|\Psi\rangle$, in general, is%

\begin{equation}
\pi_{ij}\left(  x\right)  |\Psi\rangle=-i\frac{\delta}{\delta h_{ij}\left(
\overrightarrow{x}\right)  }\Psi\left[  h_{ij}\right]  ,
\end{equation}
while the inner product is defined by the functional integration:
\begin{equation}
\left\langle \Psi_{1}\mid\Psi_{2}\right\rangle =\int\left[  \mathcal{D}%
h_{ij}\right]  \Psi_{1}^{\ast}\left[  h_{ij}\right]  \Psi_{2}\left[
h_{kl}\right]  .
\end{equation}
We demand that
\begin{equation}
\frac{1}{V}\frac{\left\langle \Psi\left\vert \int_{\Sigma}d^{3}x\hat{\Lambda
}_{\Sigma,f\left(  R\right)  }^{\left(  2\right)  }\right\vert \Psi
\right\rangle }{\left\langle \Psi|\Psi\right\rangle }=\frac{1}{V}\frac
{\int\mathcal{D}\left[  g_{ij}\right]  \Psi^{\ast}\left[  h_{ij}\right]
\int_{\Sigma}d^{3}x\hat{\Lambda}_{\Sigma,f\left(  R\right)  }^{\left(
2\right)  }\Psi\left[  h_{ij}\right]  }{\int\mathcal{D}\left[  g_{ij}\right]
\Psi^{\ast}\left[  h_{ij}\right]  \Psi\left[  h_{ij}\right]  } \label{vareq}%
\end{equation}
be stationary against arbitrary variations of $\Psi\left[  h_{ij}\right]  $.
Note that Eq.$\left(  \ref{vareq}\right)  $ can be considered as the
variational analog of a Sturm-Liouville problem with the cosmological constant
regarded as the associated eigenvalue. Therefore the solution of Eq.$\left(
\ref{WDW2}\right)  $ corresponds to the minimum of Eq.$\left(  \ref{vareq}%
\right)  $. The form of $\left\langle \Psi\left\vert \hat{\Lambda}_{\Sigma
}\right\vert \Psi\right\rangle $ can be computed with the help of the wave
functional $\left(  \ref{tt}\right)  $ and with the help of%
\begin{equation}
\frac{\left\langle \Psi\left\vert h_{ij}\left(  \overrightarrow{x}\right)
\right\vert \Psi\right\rangle }{\left\langle \Psi|\Psi\right\rangle }=0
\end{equation}
and%
\begin{equation}
\frac{\left\langle \Psi\left\vert h_{ij}\left(  \overrightarrow{x}\right)
h_{kl}\left(  \overrightarrow{y}\right)  \right\vert \Psi\right\rangle
}{\left\langle \Psi|\Psi\right\rangle }=K_{ijkl}\left(  \overrightarrow
{x},\overrightarrow{y}\right)  .
\end{equation}
Extracting the TT tensor contribution, we get%
\begin{equation}
\hat{\Lambda}_{\Sigma,f\left(  R\right)  }^{\left(  2\right)  ,\bot}=\frac
{1}{4V}\int_{\Sigma}d^{3}x\sqrt{\bar{g}}G^{ijkl}\left[  \left(  2\kappa
\right)  h\left(  R\right)  K^{-1\bot}\left(  x,x\right)  _{ijkl} +\frac
{1}{\left(  2\kappa\right)  }\left(  \triangle_{2}\right)  _{j}^{a}K^{\bot
}\left(  x,x\right)  _{iakl}\right]  . \label{p22}%
\end{equation}
The propagator $K^{\bot}\left(  x,x\right)  _{iakl}$ can be represented as
\begin{equation}
K^{\bot}\left(  \overrightarrow{x},\overrightarrow{y}\right)  _{iakl}:=%
%TCIMACRO{\dsum _{\tau}}%
%BeginExpansion
{\displaystyle\sum_{\tau}}
%EndExpansion
\frac{h_{ia}^{\left(  \tau\right)  \bot}\left(  \overrightarrow{x}\right)
h_{kl}^{\left(  \tau\right)  \bot}\left(  \overrightarrow{y}\right)
}{2\lambda\left(  \tau\right)  }, \label{proptt}%
\end{equation}
where $h_{ia}^{\left(  \tau\right)  \bot}\left(  \overrightarrow{x}\right)  $
are the eigenfunctions of $\triangle_{2}$. $\tau$ denotes a complete set of
indices and $\lambda\left(  \tau\right)  $ are a set of variational parameters
to be determined by the minimization of Eq.$\left(  \ref{p22}\right)  $. The
expectation value of $\hat{\Lambda}_{\Sigma}^{\bot}$ is easily obtained by
inserting the form of the propagator into Eq.$\left(  \ref{p22}\right)  $%
\begin{equation}
-\frac{\Lambda_{c}^{\prime}\left(  \lambda_{i}\right)  }{\kappa}=\frac{1}{4}%
%TCIMACRO{\dsum _{\tau}}%
%BeginExpansion
{\displaystyle\sum_{\tau}}
%EndExpansion%
%TCIMACRO{\dsum _{i=1}^{2}}%
%BeginExpansion
{\displaystyle\sum_{i=1}^{2}}
%EndExpansion
\left[  \left(  2\kappa\right)  h\left(  R\right)  \lambda_{i}\left(
\tau\right)  +\frac{\omega_{i}^{2}\left(  \tau\right)  }{\left(
2\kappa\right)  \lambda_{i}\left(  \tau\right)  }\right]  .
\end{equation}
By minimizing with respect to the variational function $\lambda_{i}\left(
\tau\right)  $, we obtain the total one loop energy density for TT tensors%
\begin{equation}
\Lambda_{c}^{\prime}\left(  \lambda_{i}\right)  =-\kappa\sqrt{h\left(
R\right)  }\frac{1}{4}%
%TCIMACRO{\dsum _{\tau}}%
%BeginExpansion
{\displaystyle\sum_{\tau}}
%EndExpansion
\left[  \sqrt{\omega_{1}^{2}\left(  \tau\right)  }+\sqrt{\omega_{2}^{2}\left(
\tau\right)  }\right]  , \label{lambda1loop}%
\end{equation}
where $\Lambda_{c}^{\prime}$ is expressed by the Eq.$\left(  \ref{NewLambda}%
\right)  $. The above expression makes sense only for $\omega_{i}^{2}\left(
\tau\right)  >0$. It is the main formal result of this paper. It is true for
generic $f(R)$ functions since $h(R)$ explicitly appears in it.

\section{The transverse traceless (TT) spin 2 operator for the Schwarzschild
metric and the WKB approximation}

The above considerations can be specified choosing a given metric.
For example, the quantity $\Lambda_{c}^{\prime}$ can be calculated
for a Schwarzschild metric in the WKB approximation. Apparently,
there is no a strong motivation to consider a Schwarzschild metric
as a probe for a cosmological problem. Nevertheless, every quantum
field induces a \textquotedblleft cosmological
term\textquotedblright\ by means of vacuum expectation values and
the variational approach we have considered is particularly easy
to use for a spherically symmetric metric. The Schwarzschild
metric is the simplest sourceless solution of the Einstein field
equations which can be used to compute a cosmological constant
spectrum. Of course, also Minkowski space can be put in the form
of a spherically symmetric metric, but in that case there is no
gravity at all. The other solution need a source which is not
considered in the present paper. In this sense, the computation is
a real vacuum contribution to the cosmological term. The spin-two
operator
for the Schwarzschild metric is defined by%
\begin{equation}
\left(  \triangle_{2}h^{TT}\right)  _{i}^{j}:=-\left(  \triangle_{T}%
h^{TT}\right)  _{i}^{j}+2\left(  Rh^{TT}\right)  _{i}^{j},\label{spin2}%
\end{equation}
where the transverse-traceless (TT) tensor for the quantum fluctuation is
obtained by the following decomposition%
\begin{equation}
h_{i}^{j}=h_{i}^{j}-\frac{1}{3}\delta_{i}^{j}h+\frac{1}{3}\delta_{i}%
^{j}h=\left(  h^{T}\right)  _{i}^{j}+\frac{1}{3}\delta_{i}^{j}h.
\end{equation}
This implies that $\left(  h^{T}\right)  _{i}^{j}\delta_{j}^{i}=0$. The
transversality condition is applied on $\left(  h^{T}\right)  _{i}^{j}$ and
becomes $\nabla_{j}\left(  h^{T}\right)  _{i}^{j}=0$. Thus%
\begin{equation}
-\left(  \triangle_{T}h^{TT}\right)  _{i}^{j}=-\triangle_{S}\left(
h^{TT}\right)  _{i}^{j}+\frac{6}{r^{2}}\left(  1-\frac{2MG}{r}\right)
,\label{tlap}%
\end{equation}
where $\triangle_{S}$ is the scalar curved Laplacian, whose form is%
\begin{equation}
\triangle_{S}=\left(  1-\frac{2MG}{r}\right)  \frac{d^{2}}{dr^{2}}+\left(
\frac{2r-3MG}{r^{2}}\right)  \frac{d}{dr}-\frac{L^{2}}{r^{2}}\label{slap}%
\end{equation}
and $R_{j\text{ }}^{a}$ is the mixed Ricci tensor whose components are:
\begin{equation}
R_{i}^{a}=\left\{  -\frac{2MG}{r^{3}},\frac{MG}{r^{3}},\frac{MG}{r^{3}%
}\right\}  ,
\end{equation}
This implies that the scalar curvature is traceless. We are therefore led to
study the following eigenvalue equation
\begin{equation}
\left(  \triangle_{2}h^{TT}\right)  _{i}^{j}=\omega^{2}h_{j}^{i}\label{p31}%
\end{equation}
where $\omega^{2}$ is the eigenvalue of the corresponding equation. In doing
so, we follow Regge and Wheeler in analyzing the equation as modes of definite
frequency, angular momentum and parity\cite{Regge Wheeler}. In particular, our
choice for the three-dimensional gravitational perturbation is represented by
its even-parity form%
\begin{equation}
\left(  h^{even}\right)  _{j}^{i}\left(  r,\vartheta,\phi\right)  =diag\left[
H\left(  r\right)  ,K\left(  r\right)  ,L\left(  r\right)  \right]
Y_{lm}\left(  \vartheta,\phi\right)  ,\label{pert}%
\end{equation}
with%
\begin{equation}
\left\{
\begin{array}
[c]{c}%
H\left(  r\right)  =h_{1}^{1}\left(  r\right)  -\frac{1}{3}h\left(  r\right)
\\
K\left(  r\right)  =h_{2}^{2}\left(  r\right)  -\frac{1}{3}h\left(  r\right)
\\
L\left(  r\right)  =h_{3}^{3}\left(  r\right)  -\frac{1}{3}h\left(  r\right)
\end{array}
\right.  .
\end{equation}
From the transversality condition, we obtain $h_{2}^{2}\left(  r\right)
=h_{3}^{3}\left(  r\right)  $. Then $K\left(  r\right)  =L\left(  r\right)  $.
For a generic value of the angular momentum $L$, representation $\left(
\ref{pert}\right)  $ joined to Eq.$\left(  \ref{tlap}\right)  $ lead to the
following system of PDE's%

\begin{equation}
\left\{
\begin{array}
[c]{c}%
\left(  -\triangle_{S}+\frac{6}{r^{2}}\left(  1-\frac{2MG}{r}\right)
-\frac{4MG}{r^{3}}\right)  H\left(  r\right)  =\omega_{1,l}^{2}H\left(
r\right) \\
\\
\left(  -\triangle_{S}+\frac{6}{r^{2}}\left(  1-\frac{2MG}{r}\right)
+\frac{2MG}{r^{3}}\right)  K\left(  r\right)  =\omega_{2,l}^{2}K\left(
r\right)
\end{array}
\right.  . \label{p33}%
\end{equation}
Defining the "reduced" fields%

\begin{equation}
H\left(  r\right)  =\frac{f_{1}\left(  r\right)  }{r};\qquad K\left(
r\right)  =\frac{f_{2}\left(  r\right)  }{r},
\end{equation}
and passing to the proper geodesic distance from the \textit{throat} of the
bridge%
\begin{equation}
dx=\pm\frac{dr}{\sqrt{1-\frac{2MG}{r}}}, \label{throat}%
\end{equation}
the system $\left(  \ref{p33}\right)  $ becomes%

\begin{equation}
\left\{
\begin{array}
[c]{c}%
\left[  -\frac{d^{2}}{dx^{2}}+V_{1}\left(  r\right)  \right]  f_{1}\left(
x\right)  =\omega_{1,l}^{2}f_{1}\left(  x\right) \\
\\
\left[  -\frac{d^{2}}{dx^{2}}+V_{2}\left(  r\right)  \right]  f_{2}\left(
x\right)  =\omega_{2,l}^{2}f_{2}\left(  x\right)
\end{array}
\right.  \label{p34}%
\end{equation}
with
\begin{equation}
\left\{
\begin{array}
[c]{c}%
V_{1}\left(  r\right)  =\frac{l\left(  l+1\right)  }{r^{2}}+U_{1}\left(
r\right)  +m_{g}^{2}\\
\\
V_{2}\left(  r\right)  =\frac{l\left(  l+1\right)  }{r^{2}}+U_{2}\left(
r\right)  +m_{g}^{2}%
\end{array}
\right.  ,
\end{equation}
where we have defined $r\equiv r\left(  x\right)  $ and%
\begin{equation}
\left\{
\begin{array}
[c]{c}%
U_{1}\left(  r\right)  =\left[  \frac{6}{r^{2}}\left(  1-\frac{2MG}{r}\right)
-\frac{3MG}{r^{3}}\right] \\
\\
U_{2}\left(  r\right)  =\left[  \frac{6}{r^{2}}\left(  1-\frac{2MG}{r}\right)
+\frac{3MG}{r^{3}}\right]
\end{array}
\right.  .
\end{equation}
Note that%
\begin{equation}
\left\{
\begin{array}
[c]{c}%
U_{1}\left(  r\right)  \geq0\qquad\text{when }r\geq\frac{5MG}{2}\\
U_{1}\left(  r\right)  <0\qquad\text{when }2MG\leq r<\frac{5MG}{2}\\
\\
U_{2}\left(  r\right)  >0\text{ }\forall r\in\left[  2MG,+\infty\right)
\end{array}
\right.  . \label{negU}%
\end{equation}
In order to use the WKB approximation, we define two r-dependent radial wave
numbers $k_{1}\left(  r,l,\omega_{1,nl}\right)  $ and $k_{2}\left(
r,l,\omega_{2,nl}\right)  $%
\begin{equation}
\left\{
\begin{array}
[c]{c}%
k_{1}^{2}\left(  r,l,\omega_{1,nl}\right)  =\omega_{1,nl}^{2}-\frac{l\left(
l+1\right)  }{r^{2}}-m_{1}^{2}\left(  r\right) \\
\\
k_{2}^{2}\left(  r,l,\omega_{2,nl}\right)  =\omega_{2,nl}^{2}-\frac{l\left(
l+1\right)  }{r^{2}}-m_{2}^{2}\left(  r\right)
\end{array}
\right.  , \label{rwn}%
\end{equation}
where we have defined two r-dependent effective masses $m_{1}^{2}\left(
r\right)  $ and $m_{2}^{2}\left(  r\right)  $. The WKB approximation we will
use to evaluate Eq.$\left(  \ref{lambda1loop}\right)  $ is equivalent to the
scattering phase shift method and to the entropy computation in the brick wall
model. We begin by counting the number of modes with frequency less than
$\omega_{i}$, $i=1,2$. This is given approximately by%
\begin{equation}
\tilde{g}\left(  \omega_{i}\right)  =\int\nu_{i}\left(  l,\omega_{i}\right)
\left(  2l+1\right)  , \label{p41}%
\end{equation}
where $\nu_{i}\left(  l,\omega_{i}\right)  $, $i=1,2$ is the number of nodes
in the mode with $\left(  l,\omega_{i}\right)  $, such that $\left(  r\equiv
r\left(  x\right)  \right)  $
\begin{equation}
\nu_{i}\left(  l,\omega_{i}\right)  =\frac{1}{2\pi}\int_{-\infty}^{+\infty
}dx\sqrt{k_{i}^{2}\left(  r,l,\omega_{i}\right)  }. \label{p42}%
\end{equation}
Here it is understood that the integration with respect to $x$ and $l$ is
taken over those values which satisfy $k_{i}^{2}\left(  r,l,\omega_{i}\right)
\geq0,$ $i=1,2$. With the help of Eqs.$\left(  \ref{p41},\ref{p42}\right)  $,
we obtain the one loop total energy for TT tensors which is%
\begin{equation}
\frac{1}{8\pi}\sum_{i=1}^{2}\int_{-\infty}^{+\infty}dx\left[  \int
_{0}^{+\infty}\omega_{i}\frac{d\tilde{g}\left(  \omega_{i}\right)  }%
{d\omega_{i}}d\omega_{i}\right]  .
\end{equation}
By extracting the energy density contributing to the cosmological constant, we
get%
\begin{equation}
\Lambda_{c}^{\prime}=\Lambda_{c,1}^{\prime}+\Lambda_{c,2}^{\prime}=\rho
_{1}+\rho_{2} =-\sqrt{h\left(  R\right)  }\frac{\kappa}{16\pi^{2}}\left\{
\int_{0}^{+\infty}\omega_{1}^{2}\sqrt{\omega_{1}^{2}-m_{1}^{2}\left(
r\right)  }d\omega_{1}+\int_{0}^{+\infty}\omega_{2}^{2}\sqrt{\omega_{2}%
^{2}-m_{2}^{2}\left(  r\right)  }d\omega_{2}\right\}  , \label{tote1loop}%
\end{equation}
where we have included an additional $4\pi$ coming from the angular integration.

\section{One-loop energy Regularization and Renormalization}

In this section, we will use the zeta function regularization method to
compute the energy densities $\rho_{1}$ and $\rho_{2}$. Note that this
procedure is completely equivalent to the subtraction procedure of the Casimir
energy computation where the zero point energy (ZPE) in different backgrounds
with the same asymptotic properties is involved. To this purpose, we introduce
the additional mass parameter $\mu$ in order to restore the correct dimension
for the regularized quantities. Such an arbitrary mass scale emerges
unavoidably in any regularization scheme. Then we have%
\begin{equation}
\rho_{i}\left(  \varepsilon\right)  =-\sqrt{h\left(  R\right)  }\frac{\kappa
}{16\pi^{2}}\mu^{2\varepsilon}\int_{0}^{+\infty}d\omega_{i}\frac{\omega
_{i}^{2}}{\left(  \omega_{i}^{2}-m_{i}^{2}\left(  r\right)  \right)
^{\varepsilon-\frac{1}{2}}},\label{zeta}%
\end{equation}
where%
\begin{equation}
\left\{
\begin{array}
[c]{c}%
\rho_{1}\left(  \varepsilon\right)  =-\sqrt{h\left(  R\right)  }\frac{\kappa
}{16\pi^{2}}\int_{0}^{+\infty}\omega_{1}^{2}\sqrt{\omega_{1}^{2}-m_{1}%
^{2}\left(  r\right)  }d\omega_{1}\\
\\
\rho_{2}\left(  \varepsilon\right)  =-\sqrt{h\left(  R\right)  }\frac{\kappa
}{16\pi^{2}}\int_{0}^{+\infty}\omega_{2}^{2}\sqrt{\omega_{2}^{2}-m_{2}%
^{2}\left(  r\right)  }d\omega_{2}%
\end{array}
\right.  .\label{edens}%
\end{equation}
The integration has to be meant in the range where $\omega_{i}^{2}-m_{i}%
^{2}\left(  r\right)  \geq0$\footnote{Details of the calculation can be found
in the Appendix A.}. One gets%
\begin{equation}
\rho_{i}\left(  \varepsilon\right)  =\sqrt{h\left(  R\right)  }\kappa
\frac{m_{i}^{4}\left(  r\right)  }{256\pi^{2}}\left[  \frac{1}{\varepsilon
}+\ln\left(  \frac{\mu^{2}}{m_{i}^{2}\left(  r\right)  }\right)  +2\ln
2-\frac{1}{2}\right]  ,\label{zeta1}%
\end{equation}
$i=1,2$. In order to renormalize the divergent ZPE, we write%
\begin{equation}
\Lambda_{c}^{\prime}=8\pi G\left[  \rho_{1}\left(  \varepsilon\right)
+\rho_{2}\left(  \varepsilon\right)  +\rho_{1}\left(  \mu\right)  +\rho
_{2}\left(  \mu\right)  \right]  ,
\end{equation}
where we have separated the divergent part from the finite part. For practical
purposes, it is useful to divide $\Lambda_{c}^{\prime}$ with the factor
$\sqrt{h\left(  R\right)  }$. To this aim, we define%
\begin{equation}
\frac{\Lambda_{c}^{\prime}}{\sqrt{h\left(  R\right)  }}=\left[  \Lambda
_{c}+\frac{1}{2V}\int_{\Sigma}d^{3}x\sqrt{g}\frac{Rf^{\prime}\left(  R\right)
-f\left(  R\right)  }{f^{\prime}\left(  R\right)  }\right]  \frac{1}%
{\sqrt{h\left(  R\right)  }}%
\end{equation}
and we extract the divergent part of $\Lambda$, in the limit $\varepsilon
\rightarrow0$, by setting%
\begin{equation}
\Lambda^{div}=8\pi G\left[  \rho_{1}\left(  \varepsilon\right)  +\rho
_{2}\left(  \varepsilon\right)  \right]  =\frac{G}{32\pi\varepsilon}\left[
m_{1}^{4}\left(  r\right)  +m_{2}^{4}\left(  r\right)  \right]  .
\end{equation}
Thus, the renormalization is performed via the absorption of the divergent
part into the re-definition of the bare classical cosmological constant
$\Lambda_{c}$, that is
\begin{equation}
\Lambda_{c}\rightarrow\Lambda_{0}+\sqrt{h\left(  R\right)  }\Lambda^{div}.
\end{equation}
The remaining finite value for the cosmological constant reads\footnote{Since
$m_{1}^{2}\left(  r\right)  $ can change in sign, when we integrate over
$\omega_{1}$ we can use either $I_{+}$ or $I_{-}$. This leads to the
appearance of the absolute value.}%
\[
\frac{\Lambda_{0}^{\prime}\left(  \mu\right)  }{8\pi G}=\rho_{1}\left(
\mu\right)  +\rho_{2}\left(  \mu\right)  =\frac{1}{256\pi^{2}}\left\{
m_{1}^{4}\left(  r\right)  \left[  \ln\left(  \frac{\mu^{2}}{\left\vert
m_{1}^{2}\left(  r\right)  \right\vert }\right)  +2\ln2-\frac{1}{2}\right]
+\right.
\]%
\begin{equation}
\left.  +m_{2}^{4}\left(  r\right)  \left[  \ln\left(  \frac{\mu^{2}}%
{m_{2}^{2}\left(  r\right)  }\right)  +2\ln2-\frac{1}{2}\right]  \right\}
=\rho_{eff}^{TT}\left(  \mu,r\right)  ,\label{lambda0}%
\end{equation}
where%
\begin{equation}
\Lambda_{0}^{\prime}\left(  \mu\right)  =\frac{1}{\sqrt{h\left(  R\right)  }%
}\left[  \Lambda_{0}\left(  \mu\right)  +\frac{1}{2V}\int_{\Sigma}d^{3}%
x\sqrt{g}\frac{Rf^{\prime}\left(  R\right)  -f\left(  R\right)  }{f^{\prime
}\left(  R\right)  }\right]
\end{equation}
is the modified cosmological constant. The quantity in Eq.$\left(
\ref{lambda0}\right)  $ depends on the arbitrary mass scale $\mu.$ It is
appropriate to use the renormalization group equation to eliminate such a
dependence. To this aim, we impose that \cite{RGeq}%
\begin{equation}
\frac{1}{8\pi G}\mu\frac{\partial\Lambda_{0}^{\prime}\left(  \mu\right)
}{\partial\mu}=\mu\frac{d}{d\mu}\rho_{eff}^{TT}\left(  \mu,r\right)
.\label{rg}%
\end{equation}
Solving it, we find that the renormalized constant $\Lambda_{0}$ should be
treated as a running one in the sense that it varies, provided that the scale
$\mu$ is changing
\begin{equation}
\Lambda_{0}^{\prime}\left(  \mu,r\right)  =\Lambda_{0}^{\prime}\left(  \mu
_{0},r\right)  +\frac{G}{16\pi}\left[  m_{1}^{4}\left(  r\right)  +m_{2}%
^{4}\left(  r\right)  \right]  \ln\frac{\mu}{\mu_{0}}.\label{lambdamu}%
\end{equation}
Substituting Eq.$\left(  \ref{lambdamu}\right)  $ into Eq.$\left(
\ref{lambda0}\right)  $ we find%
\begin{equation}
\frac{\Lambda_{0}^{\prime}\left(  \mu_{0},r\right)  }{8\pi G}=-\frac{1}%
{256\pi^{2}}\left\{  m_{1}^{4}\left(  r\right)  \left[  \ln\left(
\frac{\left\vert m_{1}^{2}\left(  r\right)  \right\vert }{\mu_{0}^{2}}\right)
-2\ln2+\frac{1}{2}\right]  +m_{2}^{4}\left(  r\right)  \left[  \ln\left(
\frac{m_{2}^{2}\left(  r\right)  }{\mu_{0}^{2}}\right)  -2\ln2+\frac{1}%
{2}\right]  \right\}  .\label{lambdamu0}%
\end{equation}
It is worth remarking that while $m_{2}^{2}\left(  r\right)  $ is constant in
sign, $m_{1}^{2}\left(  r\right)  $ is not. Indeed, for the critical value
$\bar{r}=5MG/2$, $m_{1}^{2}\left(  \bar{r}\right)  =m_{g}^{2}$ and in the
range $\left(  2MG,5MG/2\right)  $ for some values of $m_{g}^{2}$, $m_{1}%
^{2}\left(  \bar{r}\right)  $ can be negative. It is interesting therefore
concentrate in this range. To further proceed, we observe that $m_{1}%
^{2}\left(  r\right)  $ and $m_{2}^{2}\left(  r\right)  $ can be recast into a
more suggestive and useful form, namely%
\begin{equation}
\left\{
\begin{array}
[c]{c}%
m_{1}^{2}\left(  r\right)  =U_{1}\left(  r\right)  =m_{1}^{2}\left(
r,M\right)  -m_{2}^{2}\left(  r,M\right)  \\
\\
m_{2}^{2}\left(  r\right)  =U_{2}\left(  r\right)  =m_{1}^{2}\left(
r,M\right)  +m_{2}^{2}\left(  r,M\right)
\end{array}
\right.  ,
\end{equation}
where $m_{1}^{2}\left(  r,M\right)  \rightarrow0$ when $r\rightarrow\infty$ or
$r\rightarrow2MG$ and $m_{2}^{2}\left(  r,M\right)  =3MG/r^{3}$. Nevertheless,
in the above mentioned range $m_{1}^{2}\left(  r,M\right)  $ is negligible
when compared with $m_{2}^{2}\left(  r,M\right)  $. So, in a first
approximation we can write%
\begin{equation}
\left\{
\begin{array}
[c]{c}%
m_{1}^{2}\left(  r\right)  \simeq-m_{2}^{2}\left(  r_{0},M\right)  =-m_{0}%
^{2}\left(  M\right)  \\
\\
m_{2}^{2}\left(  r\right)  \simeq m_{2}^{2}\left(  r_{0},M\right)  =m_{0}%
^{2}\left(  M\right)
\end{array}
\right.  ,
\end{equation}
where we have defined a parameter $r_{0}>2MG$ and $m_{0}^{2}\left(  M\right)
=3MG/r_{0}^{3}$. The main reason for introducing a new parameter resides in
the fluctuation of the horizon that forbids any kind of approach. Of course
the quantum fluctuation must obey the uncertainty relations. Thus Eq.$\left(
\ref{lambdamu0}\right)  $ becomes%
\begin{equation}
\frac{\Lambda_{0}^{\prime}\left(  \mu_{0},r\right)  }{8\pi G}=-\frac{m_{0}%
^{4}\left(  M\right)  }{128\pi^{2}}\left[  \ln\left(  \frac{m_{0}^{2}\left(
M\right)  }{4\mu_{0}^{2}}\right)  +\frac{1}{2}\right]  .\label{lambdamu0a}%
\end{equation}
Now, we compute the maximum of $\Lambda_{0}^{\prime}$, by setting%
\begin{equation}
x=\frac{m_{0}^{2}\left(  M\right)  }{4\mu_{0}^{2}}.
\end{equation}
Thus $\Lambda_{0}^{\prime}$ becomes%
\begin{equation}
\Lambda_{0}^{\prime}\left(  \mu_{0},x\right)  =-\frac{G\mu_{0}^{4}}{\pi}%
x^{2}\left[  \ln\left(  x\right)  +\frac{1}{2}\right]  .\label{LambdansM}%
\end{equation}
As a function of $x$, $\Lambda_{0}\left(  \mu_{0},x\right)  $ vanishes for
$x=0$ and $x=\exp\left(  -\frac{1}{2}\right)  $ and when $x\in\left[
0,\exp\left(  -\frac{1}{2}\right)  \right]  $, $\Lambda_{0}^{\prime}\left(
\mu_{0},x\right)  \geq0$. It has a maximum for
\begin{equation}
\bar{x}=\frac{1}{e}\qquad\Longleftrightarrow\qquad m_{0}^{2}\left(  M\right)
=\frac{4\mu_{0}^{2}}{e}%
\end{equation}
and its value is%
\begin{equation}
\Lambda_{0}^{\prime}\left(  \mu_{0},\bar{x}\right)  =\frac{G\mu_{0}^{4}}{2\pi
e^{2}}\qquad\text{or\qquad}\frac{1}{\sqrt{h\left(  R\right)  }}\left[
\Lambda_{0}\left(  \mu_{0},\bar{x}\right)  +\frac{1}{2V}\int_{\Sigma}%
d^{3}x\sqrt{g}\frac{Rf^{\prime}\left(  R\right)  -f\left(  R\right)
}{f^{\prime}\left(  R\right)  }\right]  =\frac{G\mu_{0}^{4}}{2\pi e^{2}}.
\end{equation}
Isolating $\Lambda_{0}\left(  \mu_{0},\bar{x}\right)  $, we get%
\begin{equation}
\Lambda_{0}\left(  \mu_{0},\bar{x}\right)  =\sqrt{h\left(  R\right)  }%
\frac{G\mu_{0}^{4}}{2\pi e^{2}}-\frac{1}{2V}\int_{\Sigma}d^{3}x\sqrt{g}%
\frac{Rf^{\prime}\left(  R\right)  -f\left(  R\right)  }{f^{\prime}\left(
R\right)  }.
\end{equation}
Note that $\Lambda_{0}\left(  \mu_{0},\bar{x}\right)  $ can be set to zero
when%
\begin{equation}
\sqrt{h\left(  R\right)  }\frac{G\mu_{0}^{4}}{2\pi e^{2}}=\frac{1}{2V}%
\int_{\Sigma}d^{3}x\sqrt{g}\frac{Rf^{\prime}\left(  R\right)  -f\left(
R\right)  }{f^{\prime}\left(  R\right)  }.\label{lambda0_fin}%
\end{equation}
Let us see what happens when $f\left(  R\right)  =\exp\left(
-\alpha R\right)  $. This choice is simply suggested by the
regularity of the function at every scale \textbf{and by the fact
that any power of $R$, considered as a correction to GR, is
included}. In this case, Eq.$\left( \ref{lambda0_fin}\right)
$ becomes%
\begin{equation}
\sqrt{\frac{3\alpha\exp\left(  -\alpha R\right)  +2}{\alpha\exp\left(  -\alpha
R\right)  }}\frac{G\mu_{0}^{4}}{\pi e^{2}}=\frac{1}{\alpha V}\int_{\Sigma
}d^{3}x\sqrt{g}\left(  1+\alpha R\right)  .
\end{equation}
For Schwarzschild, it is $R=0$, then%
\begin{equation}
\frac{G\mu_{0}^{4}}{\pi e^{2}}=\sqrt{\frac{1}{\left(  3\alpha+2\right)
\alpha}}.
\end{equation}
By setting $\alpha=G$, we have the relation%
\begin{equation}
\mu_{0}^{4}=\frac{\pi e^{2}}{G}\sqrt{\frac{1}{\left(  3G+2\right)  G}}.
\end{equation}

\textbf{Remark }Note that in any case, the maximum of $\Lambda$ corresponds to
the minimum of the energy density.

\section{Summary and Conclusions}

Despite of the successes of General Relativity, such a theory can
only be considered as a step toward a much more complete and
comprehensive structure due to a large number of weaknesses. Among
them, the issue to find out the fundamental gravitational vacuum
state is one of the main problem to achieve a definite Quantum
Gravity theory which, till now is lacking. However, several
semiclassical approaches have been proposed and, from several
points of view, it is clear that the former Hilbert-Einstein
scheme has to be enlarged. The $f(R)$ theories of gravity are a
minimal but well founded extension of GR where the form of the
function $f(R)$ is not supposed "a priori" but is reconstructed by
the observed dynamics at galactic and cosmological scales
\cite{mimicking,jcap}. Also if they seems a viable scheme from
cosmology and astrophysics viewpoints, their theoretical
foundation has to be sought  at a fundamental level. In
particular, one has to face the possibility to encompass the
$f(R)$ gravity in the general framework of Quantum Field Theory on
curved spacetime. In this paper, we have dealt with the problem to
find out vacuum states for $f(R)$ gravity via the $(3+1)$ ADM
formalism. Analogously to GR, we have constructed the Hamiltonian
constraint of a generic $f(R)$ theory and then achieved a
canonical quantization giving the $f(R)$-WDW equation. In this
context, the cosmological constant (vacuum state) emerges as a WDW
eigenvalue. The related wave functional can be split by an
orthogonal decomposition and then, constructing the
transverse-traceless propagator, it is possible to obtain, after a
variational minimization, the total one-loop energy density for
the TT tensors. Such a quantity explicitly depends on the form of
$f(R)$. As an application, we derive the energy density
contributions to the cosmological constant for a TT spin 2
operator in the Schwarzschild metric and in the WKB approximation.
The one-loop energy regularization and renormalization are
achieved by the zeta function regularization method. The resulting
renormalized $\Lambda_{0}$ is a running constant which can be set
to zero depending on the value of an arbitrary mass scale
parameter $\mu$. As explicit calculation, we find out the value of
such a parameter for a theory of the form $f(R)=\exp(-\alpha R)$
in the Schwarzschild metric. \textbf{This case can be used for
several applications at cosmological and astrophysical scales. In
particular, truncated versions of such an exponential function,
power law $f(R)$, have been used for galactic dynamics
\cite{noipla,jcap,mnras}. In those cases, a corrected Newtonian
potential, derived from the $f(R)$ Schwarzschild solution, has
been used to fit, with great accuracy,
 data from low surface brightness galaxies without using dark
matter haloes. This approach has allowed to fix a suitable mass
scale comparable with the core size of galactic systems. Such a
mass can be directly related to the above parameter $\alpha$
depending on the core radius $r_c$ (see also \cite{mnras}).}

In summary, the application of Quantum Field Theory methods to $f(R)$ gravity
seems a viable scheme and gives positive results toward the issue to select
vacuum states (eigenvalues) which can be interpreted as the cosmological
constant. However, further studies are needed in order to generalize such
results to other metrics and other Extended Theories of Gravity.

\appendix

\section{Zeta function regularization}

\label{app}In this appendix, we report details on computation leading to
expression $\left(  \ref{zeta}\right)  $. We begin with the following integral%
\begin{equation}
\rho\left(  \varepsilon\right)  =\left\{
\begin{array}
[c]{c}%
I_{+}=\mu^{2\varepsilon}\int_{0}^{+\infty}d\omega\frac{\omega^{2}}{\left(
\omega^{2}+m^{2}\left(  r\right)  \right)  ^{\varepsilon-\frac{1}{2}}}\\
\\
I_{-}=\mu^{2\varepsilon}\int_{0}^{+\infty}d\omega\frac{\omega^{2}}{\left(
\omega^{2}-m^{2}\left(  r\right)  \right)  ^{\varepsilon-\frac{1}{2}}}%
\end{array}
\right.  , \label{rho}%
\end{equation}
with $m^{2}\left(  r\right)  >0$.

\subsection{$I_{+}$ computation}

\label{app1}If we define $t=\omega/\sqrt{m^{2}\left(  r\right)  }$, the
integral $I_{+}$ in Eq.$\left(  \ref{rho}\right)  $ becomes%
\[
\rho\left(  \varepsilon\right)  =\mu^{2\varepsilon}m^{4-2\varepsilon}\left(
r\right)  \int_{0}^{+\infty}dt\frac{t^{2}}{\left(  t^{2}+1\right)
^{\varepsilon-\frac{1}{2}}}%
\]%
\[
=\frac{1}{2}\mu^{2\varepsilon}m^{4-2\varepsilon}\left(  r\right)  B\left(
\frac{3}{2},\varepsilon-2\right)
\]%
\[
\frac{1}{2}\mu^{2\varepsilon}m^{4-2\varepsilon}\left(  r\right)  \frac
{\Gamma\left(  \frac{3}{2}\right)  \Gamma\left(  \varepsilon-2\right)
}{\Gamma\left(  \varepsilon-\frac{1}{2}\right)  }%
\]%
\begin{equation}
=\frac{\sqrt{\pi}}{4}m^{4}\left(  r\right)  \left(  \frac{\mu^{2}}%
{m^{2}\left(  r\right)  }\right)  ^{\varepsilon}\frac{\Gamma\left(
\varepsilon-2\right)  }{\Gamma\left(  \varepsilon-\frac{1}{2}\right)  },
\end{equation}
where we have used the following identities involving the beta function%
\begin{equation}
B\left(  x,y\right)  =2\int_{0}^{+\infty}dt\frac{t^{2x-1}}{\left(
t^{2}+1\right)  ^{x+y}}\qquad\operatorname{Re}x>0,\operatorname{Re}y>0
\end{equation}
related to the gamma function by means of%
\begin{equation}
B\left(  x,y\right)  =\frac{\Gamma\left(  x\right)  \Gamma\left(  y\right)
}{\Gamma\left(  x+y\right)  }.
\end{equation}
Taking into account the following relations for the $\Gamma$-function%
\begin{equation}%
\begin{array}
[c]{c}%
\Gamma\left(  \varepsilon-2\right)  =\frac{\Gamma\left(  1+\varepsilon\right)
}{\varepsilon\left(  \varepsilon-1\right)  \left(  \varepsilon-2\right)  }\\
\\
\Gamma\left(  \varepsilon-\frac{1}{2}\right)  =\frac{\Gamma\left(
\varepsilon+\frac{1}{2}\right)  }{\varepsilon-\frac{1}{2}}%
\end{array}
, \label{gamma}%
\end{equation}
and the expansion for small $\varepsilon$%
\begin{equation}%
\begin{array}
[c]{cc}%
\Gamma\left(  1+\varepsilon\right)  = & 1-\gamma\varepsilon+O\left(
\varepsilon^{2}\right) \\
& \\
\Gamma\left(  \varepsilon+\frac{1}{2}\right)  = & \Gamma\left(  \frac{1}%
{2}\right)  -\varepsilon\Gamma\left(  \frac{1}{2}\right)  \left(  \gamma
+2\ln2\right)  +O\left(  \varepsilon^{2}\right) \\
& \\
x^{\varepsilon}= & 1+\varepsilon\ln x+O\left(  \varepsilon^{2}\right)
\end{array}
,\qquad
\end{equation}
where $\gamma$ is the Euler's constant, we find%
\begin{equation}
\rho\left(  \varepsilon\right)  =-\frac{m^{4}\left(  r\right)  }{16}\left[
\frac{1}{\varepsilon}+\ln\left(  \frac{\mu^{2}}{m^{2}\left(  r\right)
}\right)  +2\ln2-\frac{1}{2}\right]  .
\end{equation}

\subsection{$I_{-}$ computation}

\label{app2}If we define $t=\omega/\sqrt{m^{2}\left(  r\right)  }$, the
integral $I_{-}$ in Eq.$\left(  \ref{rho}\right)  $ becomes%
\[
\rho\left(  \varepsilon\right)  =\mu^{2\varepsilon}m^{4-2\varepsilon}\left(
r\right)  \int_{0}^{+\infty}dt\frac{t^{2}}{\left(  t^{2}-1\right)
^{\varepsilon-\frac{1}{2}}}%
\]%
\[
=\frac{1}{2}\mu^{2\varepsilon}m^{4-2\varepsilon}\left(  r\right)  B\left(
\varepsilon-2,\frac{3}{2}-\varepsilon\right)
\]%
\[
\frac{1}{2}\mu^{2\varepsilon}m^{4-2\varepsilon}\left(  r\right)  \frac
{\Gamma\left(  \frac{3}{2}-\varepsilon\right)  \Gamma\left(  \varepsilon
-2\right)  }{\Gamma\left(  -\frac{1}{2}\right)  }%
\]%
\begin{equation}
=-\frac{1}{4\sqrt{\pi}}m^{4}\left(  r\right)  \left(  \frac{\mu^{2}}%
{m^{2}\left(  r\right)  }\right)  ^{\varepsilon}\Gamma\left(  \frac{3}%
{2}-\varepsilon\right)  \Gamma\left(  \varepsilon-2\right)  ,
\end{equation}
where we have used the following identity involving the beta function%
\begin{equation}%
\begin{array}
[c]{c}%
\frac{1}{p}B\left(  1-\nu-\frac{\mu}{p},\nu\right)  =\int_{1}^{+\infty
}dtt^{\mu-1}\left(  t^{p}-1\right)  ^{\nu-1}\\
\\
p>0,\operatorname{Re}\nu>0,\operatorname{Re}\mu<p-p\operatorname{Re}\nu
\end{array}
\end{equation}
and the reflection formula%
\begin{equation}
\Gamma\left(  z\right)  \Gamma\left(  1-z\right)  =-z\Gamma\left(  -z\right)
\Gamma\left(  z\right)
\end{equation}
From the first of Eqs.$\left(  \ref{gamma}\right)  $ and from the expansion
for small $\varepsilon$%
\[
\Gamma\left(  \frac{3}{2}-\varepsilon\right)  =\Gamma\left(  \frac{3}%
{2}\right)  \left(  1-\varepsilon\left(  -\gamma-2\ln2+2\right)  \right)
+O\left(  \varepsilon^{2}\right)
\]%
\begin{equation}
x^{\varepsilon}=1+\varepsilon\ln x+O\left(  \varepsilon^{2}\right)  ,
\end{equation}
we find%
\begin{equation}
\rho\left(  \varepsilon\right)  =-\frac{m^{4}\left(  r\right)  }{16}\left[
\frac{1}{\varepsilon}+\ln\left(  \frac{\mu^{2}}{m^{2}\left(  r\right)
}\right)  +2\ln2-\frac{1}{2}\right]  .
\end{equation}

\end{document}